\documentclass[twocolumn,english,aps,prl,showpacs,groupedaddress]{revtex4}
\usepackage[T1]{fontenc}
\usepackage{amsmath}
\usepackage{graphicx}
\usepackage{amssymb}
\usepackage{esint}
\usepackage{xspace}
\usepackage{mydef,boldfonts}
\usepackage{subfigure}
\makeatletter

\@ifundefined{textcolor}{}
{%
 \definecolor{BLACK}{gray}{0}
 \definecolor{WHITE}{gray}{1}
 \definecolor{RED}{rgb}{1,0,0}
 \definecolor{GREEN}{rgb}{0,1,0}
 \definecolor{BLUE}{rgb}{0,0,1}
 \definecolor{CYAN}{cmyk}{1,0,0,0}
 \definecolor{MAGENTA}{cmyk}{0,1,0,0}
 \definecolor{YELLOW}{cmyk}{0,0,1,0}
 }

\usepackage{dcolumn}
\usepackage{bm}
\usepackage{epsfig}\usepackage{array}\usepackage{lscape}\@ifundefined{definecolor}
 {\usepackage{color}}{}

\def\rootfig{./}

\makeatother

\usepackage{babel}

\begin{document}

\title{Nonlinear dynamo action in a precessing cylindrical container}

\author{C. Nore$^{\text{1}}$, J. L\'eorat$^{\text{2}}$, J.-L. Guermond$^{\text{1,3}}$
and F. Luddens$^{\text{1,3}}$}

\email{nore@limsi.fr}

\affiliation{$^{\text{1}}$Laboratoire d'Informatique pour la M\'ecanique et les
Sciences de l'Ing\'enieur, CNRS UPR 3251, BP 133, 91403
Orsay cedex, France, Universit\'e Paris-Sud 11 and Institut Universitaire de France; $^{\text{2}}$Luth, Observatoire de Paris-Meudon,
place Janssen, 92195-Meudon, France; $^{\text{3}}$Department of Mathematics,
Texas A\&M University 3368 TAMU, College Station, TX 77843-3368, USA}

\date{\today}
\begin{abstract}
It is numerically demonstrated by means of a magnetohydrodynamics (MHD)
code that precession can trigger the dynamo effect in
a cylindrical container. This result adds credit to the hypothesis that
precession can be strong enough to be one of the sources of the dynamo action
in some astrophysical bodies. 
\end{abstract}


\maketitle


Simulating numerically or reproducing experimentally the dynamo action
in astrophysical bodies is challenging in many respects and the
question of converting kinetic energy into magnetic energy through a
realistic forcing mechanism is seldom addressed either numerically or
experimentally.  For instance, numerical simulations are often
restricted to \emph{kinematic dynamos}  where fluid flows are
prescribed for simplicity reasons and bear little resemblance to reality.
The three fluid dynamo experiments~\cite{Ga2000,StMu01,Monchaux07} that
have been successful so far do not model astrophysical
dynamos either, since the external pumps or internal impellers which have
been used in these experiments have no astrophysical counterparts. As
far as natural forcing mechanisms are concerned, buoyancy and
precession are believed to be possible sources of energy for the
geomagnetic dynamo. The precession hypothesis has been formulated for
the first time in \cite{malkus_precession_1968} and has since then
been actively investigated from the theoretical, experimental and
numerical perspectives \cite{Gans70,JL_PL_JLG_FP_2001,Lagrange_PoF08}.
To the best of our knowledge, however, it seems that it is only
recently that numerical examples of precession dynamos have been
reported in spheres
\cite{tilgner_precession_2005,tilgner_kinematic_2007} and in
spheroidal cavities \cite{wu_dynamo_2009}. While spheres and spheroids
are relevant for planetary dynamos, cylindrical containers seem more
convenient for experimental purpose.  In this respect we have in mind
the large scale MHD facility DRESDYN currently being built at
Helmholtz-Zentrum Dresden-Rossendorf in Germany where, among other
things, the action of precession will be tested on cylinders
(F.~Stefani, personal communication).  The objective of the present
Letter is to report numerical evidences supporting the idea that
precession is indeed a potent mechanism to drive dynamo action in
cylindrical containers.

The conducting domain considered in this letter is a cylindrical
vessel $\calC$ of radius $R$ and length $L$. The vessel contains a
conducting fluid and is embedded in vacuum.  The solid walls of the
vessel are assumed to be so thin that their influence is henceforth
neglected.  The container rotates about its axis of symmetry with
angular velocity $\Omega_r \be_z$ and is assumed to precess about a
second axis spanned by the unit vector $\be_p$ forming an angle
$\alpha$ with $\be_z$, $(0<\alpha<\pi)$. The precession angular
velocity is $\Omega_p \be_p$. A cylindrical coordinate system about
the axis of the cylinder is defined as follows: the origin of the
coordinate system is the center of mass of the cylinder, say $O$; the
$Oz$ axis is the line passing through $O$ and parallel to $\be_z$; the
origin of the angular coordinate $\theta \,(0\leq\theta\leq\pi)$ is
the half plane passing through $O$, spanned by $\be_z$ and $\be_p$,
and containing $\Omega_p \be_p$.  The third coordinate, denoted $r$,
is the distance to the $Oz$ axis.

We denote by $\calL=R$ and $\calU = R \Omega_r$ the reference length
and velocity scales, respectively. The fluid density, $\rho$, is
assumed to be constant and the reference pressure scale is
$\calP:=\rho \calU^2$. The magnetic permeability is uniform throughout
the entire space, $\mu_0$, and the electric conductivity of the
conducting fluid is constant, $\sigma_0$. The quantities $\mu_0$ and
$\sigma_0$ are used as reference magnetic permeability and electric
conductivity, respectively. The reference scale for the magnetic field
is chosen so that the reference Alfv\'en speed is $1$, \ie
$\calH:=\calU \sqrt{\rho/\mu_0}$.  We are left with five
non-dimensional parameters: one geometrical parameter $L/R$ (aspect
ratio); two forcing parameters $\alpha$ (precession angle) and
$\varepsilon=\Omega_p/\Omega_r$ (precession rate); and two fluid
parameters, namely the Ekman number $E=\nu/R^2 \Omega_r$ (where $\nu$
is the kinematic viscosity) and the magnetic Prandtl number $Pm=\nu
\mu_0 \sigma_0$.  We finally define the kinetic Reynolds number
$Re=1/E$ and the magnetic Reynolds number $Rm=Pm Re$.

The non-dimensional set of equations that we consider is written as
follows in the precessing frame of reference:
\begin{eqnarray*}
  \partial_t\bu + (\bu\ADV)\bu + 2\varepsilon \be_p \CROSS\bu + \GRAD p
  & = & \frac{1}{Re}\LAP\bu \label{eq:nsp}+ \bef, \\ \DIV \bu & = & 0,
  \label{eq:divp}\\ \partial_t\bh -\ROT(\bu \times \bh) & = &
  \frac{1}{Rm} \LAP \bh, \label{eq:ind} \\ \DIV\bh & = & 0,
\label{eq:divh}
\end{eqnarray*}
where $\bu$, $p$, and $\bh$ are the velocity field, the pressure, and
the magnetic field, respectively. In the following we consider three
different modes to solve these equations: (i) The incompressible
Navier-Stokes mode; (ii) The Maxwell or kinematic dynamo mode; (iii)
The nonlinear magnetohydrodynamics mode (MHD). In Navier-Stokes mode
the source term $\bef$ is set to zero and $\bh$ is not computed. In
Maxwell mode, only the induction equation is solved assuming that the
velocity field $\bu$ is given. In MHD mode the full set of equations
is solved and the source term $\bef$ is the Lorentz force per unit
mass, $\bef:= (\ROT\bh ) \CROSS \bh$. The no-slip boundary condition
on the velocity field is written as follows in the precessing frame of
reference: $\bu=\be_\theta$ at $r=1$ and $\bu=r\be_\theta$ at $z=\pm
1$.  The magnetic field is represented as the gradient of a scalar
potential in the vacuum, $\GRAD\phi$. The magnetic boundary
transmission conditions enforce that the magnetic field is continuous
across the walls of the vessel, say $\Sigma$, \ie $\bh|_{\Sigma}=\GRAD
\phi|_{\Sigma}$.

The above equations are solved numerically by means of a code which is
specialized to axisymmetric domains and has been presented in details
in~\cite{GLLN09,GAFD_Giesecke_2010b}. The code is called SFEMaNS for Spectral/Finite Elements for
Maxwell and Navier-Stokes equations. It is an hybrid algorithm that uses
finite element representations in the meridian section of the
axisymmetric domain and Fourier representations in the azimuthal
direction. The magnetic field is
represented as a vector field in the conducting region and as the
gradient of a scalar potential in the insulating region. SFEMaNS can
account for discontinuous distributions of electric conductivity or
magnetic permeability and all the required continuity conditions
across the interfaces are enforced using an interior penalty
technique.  The solution technique is parallel and parallelization is
done with respect to the Fourier modes.

The typical spatial resolution in the meridional plane
of the conducting domain is $\Delta x =1/160$. We take 32 Fourier
modes ($m=0,\ldots,31$) for Navier-Stokes runs and 64 Fourier
modes ($m=0,\ldots,63$) for MHD runs.  The typical time-step is
$\Delta t=0.001$. The grid is non-uniform in the vacuum with $\Delta x
=1/160$ at the cylinder walls and $\Delta x =1$ at the outer boundary
of the numerical domain, which is a sphere of radius ten times larger
than that of the cylinder. A typical MHD run requires about 1000 CPU
hours per rotation on 64 processors on an IBM-SP6.

%
%

Let us first briefly recall what is observed in a typical precessing
fluid experiment starting with the fluid at rest, (see \eg
\cite{Lagrange_PoF08,JL_PL_JLG_FP_2001}).  The vessel is first set in
rotation without precession. The fluid motion is then governed by the
formation of a viscous Ekman boundary layer during the acceleration
ramp.  The resulting flow is a stable solid rotation independently of
the strength of the acceleration phase. Once precession is applied,
the Coriolis force generates an axial motion of the flow supported by
the Fourier mode $m=1$.  When $Re$ is large enough, the flow undergoes
a transition from laminar to turbulent even for small precession rates
and small angles~\cite{Lagrange_PoF08}. The range $\varepsilon\in
[0.1,0.15]$ has been shown in \cite{JL_PL_JLG_FP_2001} to maximize the
ratio of axial to transverse energy in a cylinder of aspect ratio $2$
in the range $Rm\in [500,5000]$ when $\alpha=\pi/2$.  Although a
detailed study of the various transitions between these hydrodynamic
regimes is interesting per se, due to limited numerical resource we
reduce the dimensionality of the parametric space to one aspect ratio,
$L/R=2$, one precession angle, $\alpha=\pi/2$, one precession rate,
$\varepsilon=0.15$, and only two values of $Re\in \{1000,1200\}$ and
four values of $Rm\in \{600, 800, 1200, 2400\}$.

\begin{figure}[ht]
\centerline{
\subfigure[total kinetic energy $K$, $Re=1000$, $Re=1200$ ]{
 \includegraphics[width=0.24\textwidth]{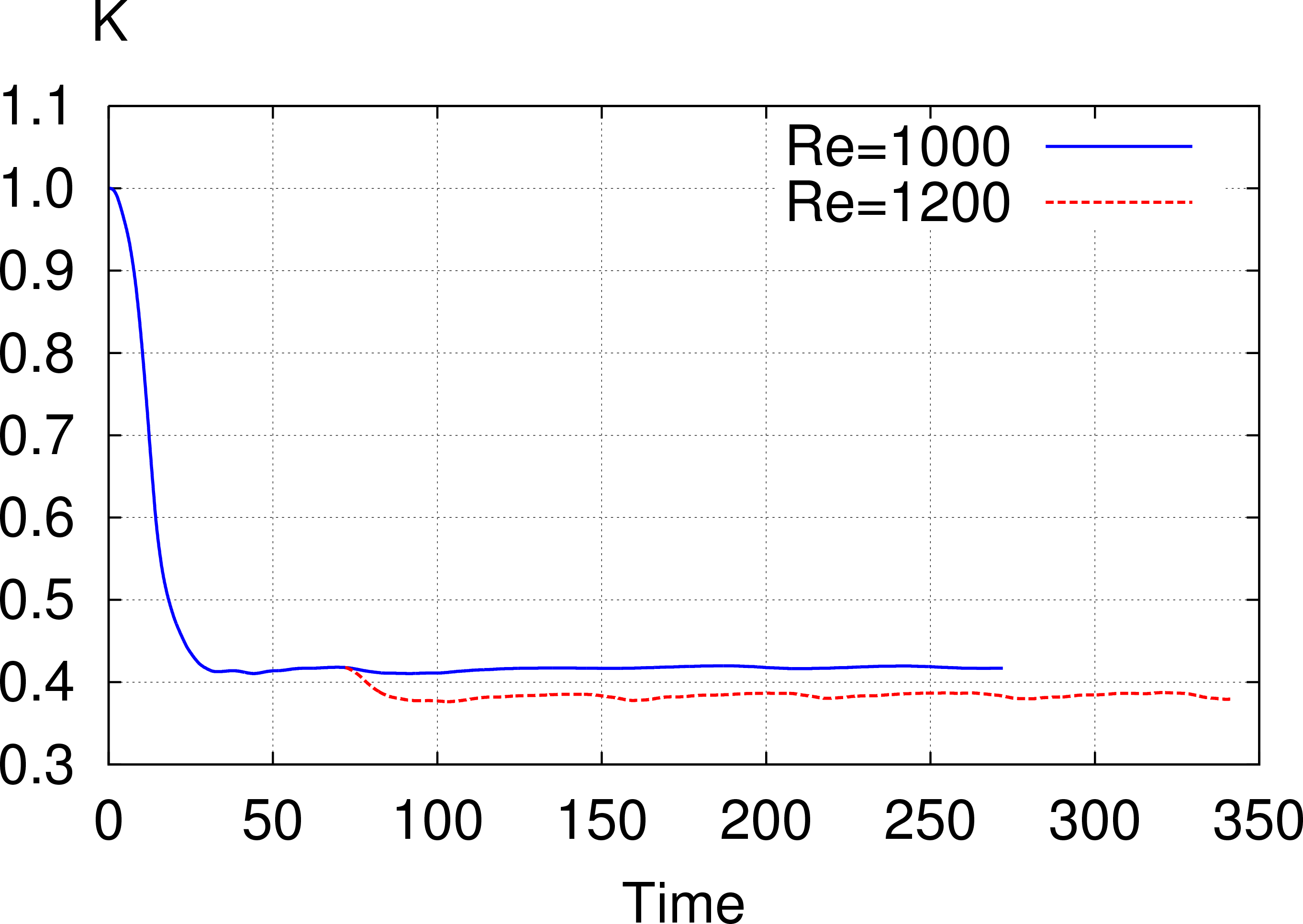}
\label{fig:K_Re1000_Re1200:a}}\hspace{-0.015\textwidth}
\subfigure[axial kinetic energy $K_z$, $Re=1000$, $Re=1200$ ]{
\includegraphics[width=0.24\textwidth]{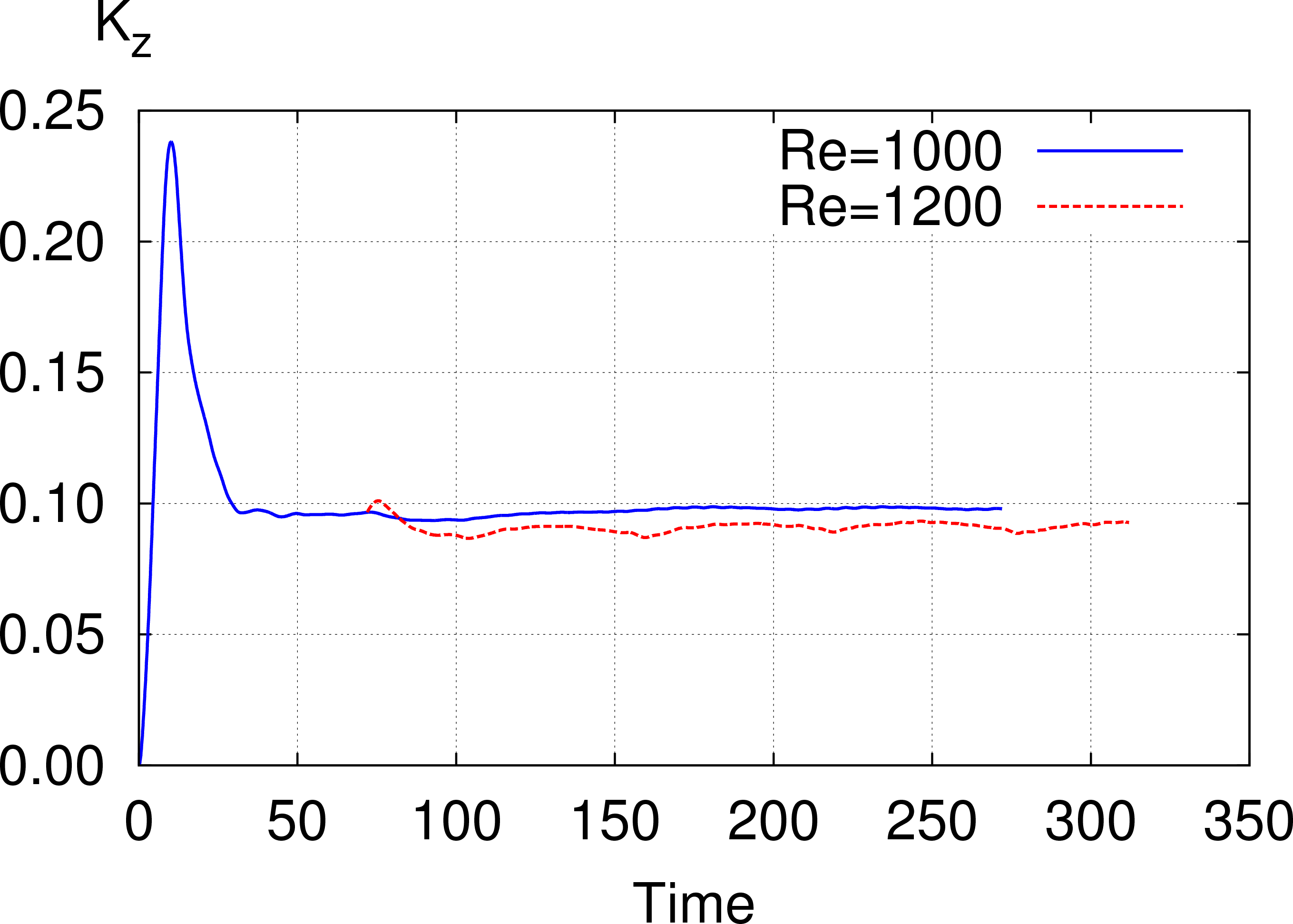}
\label{fig:K_Re1000_Re1200:b}}
}
\centerline{
\subfigure[zoom of (a), $t\in{[}50,342{]}$, $Re=1000$, $Re=1200$ ]{
\includegraphics[width=0.24\textwidth]{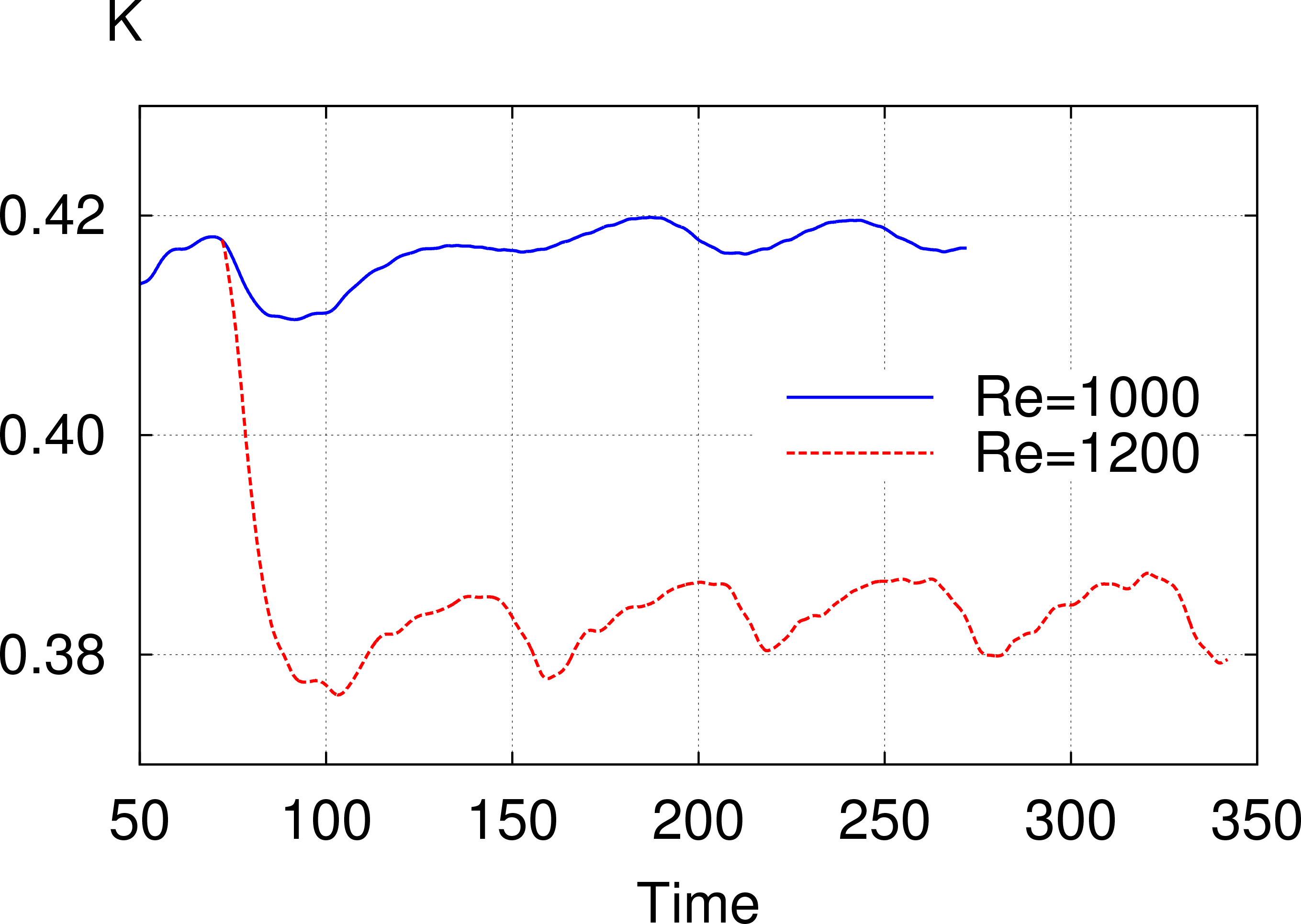}
\label{fig:K_Re1000_Re1200:c}}\hspace{-0.015\textwidth}
\subfigure[zoom of $K/2$ in (a) and north and south kinetic energies, $Re=1200$, $t\in{[}230,342{]}$]{
\includegraphics[width=0.24\textwidth]{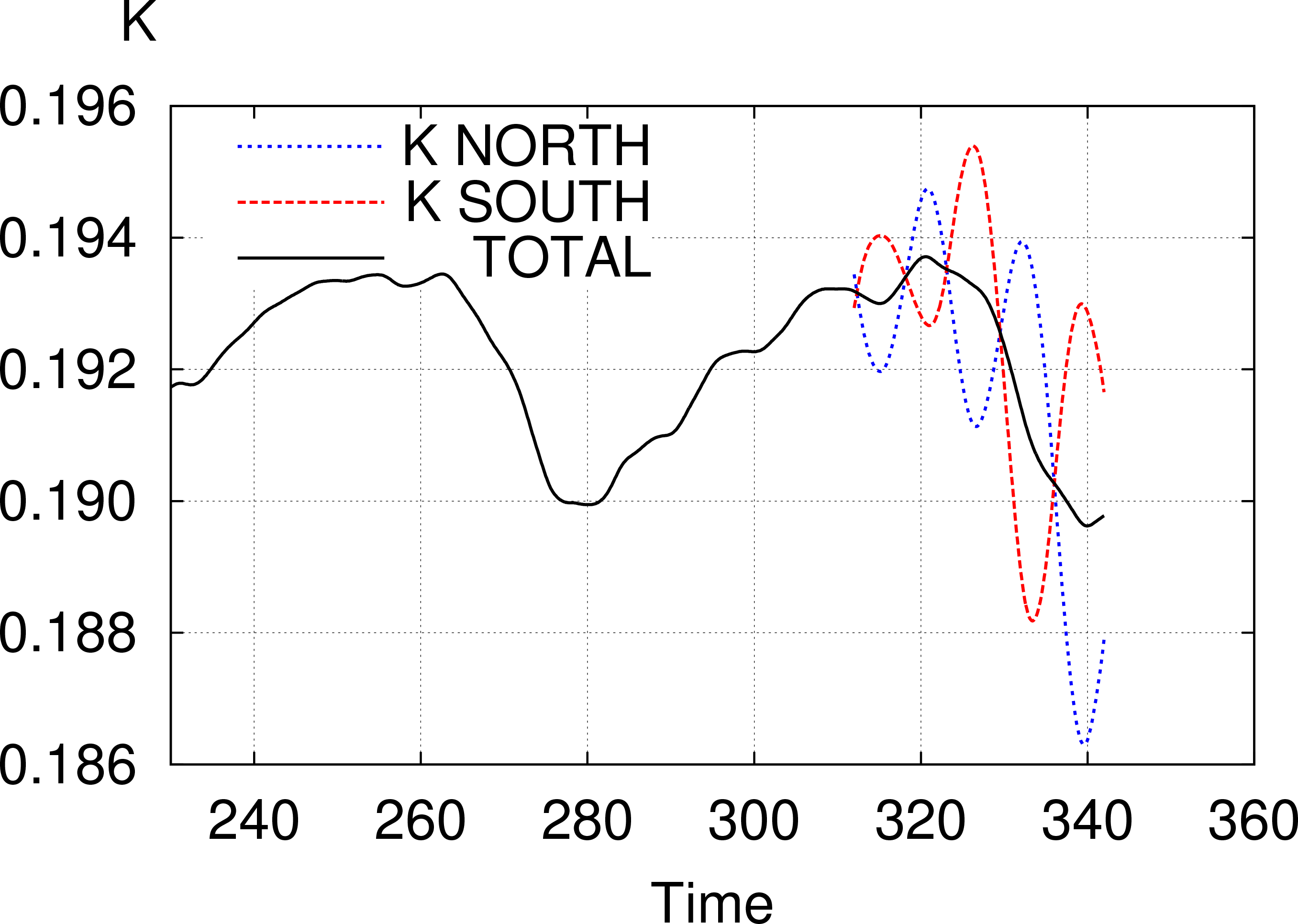}
\label{fig:K_Re1000_Re1200:d}}
}
\caption{Time evolution of the total kinetic energy $K$, axial kinetic
  energy $K_z$, and total north and south kinetic energies as
  indicated.}
\label{fig:K_Re1000_Re1200}
\end{figure}

We start our investigations with a Navier-Stokes run at $Re=1000$.
The initial velocity field is the solid rotation in the precessing
frame: $\bu_0=\be_z{\times}\br$. The onset of the axial
circulation induced by precession is monitored by recording the time
evolution of the normalized total kinetic energy $K(t)= \frac{1}{2}
\int_{\calC} \bu^{2}(\br,t) \diff\br/K_0$ and normalized axial kinetic
energy $K_z(t)= \frac{1}{2} \int_{\calC} u_z^{2}(\br,t) \diff\br/K_0$
where $K_0= \frac{1}{2} \int_{\calC} \bu_0^{2} \diff\br$ is the kinetic
energy of the initial motion. The time evolution of $K(t)$ and
$K_z(t)$ for $t\in [0,272]$ is reported in
fig.~\ref{fig:K_Re1000_Re1200}. The time $t=272$ corresponds to
$43.3$ rotation periods. After a transient that lasts
5 rotation periods and peaks at two rotation periods, the axial
kinetic energy reaches a plateau value $K_z \approx 0.1$. Meanwhile,
the total kinetic energy decreases and reaches a plateau value
$K\approx 0.42$ after 5 rotation periods also. To enrich the dynamics
of the system we have restarted the computation at $t=72$ (\ie
$11.5$ rotation periods) and increased the Reynolds number to
$Re=1200$.  The time evolution of $K(t)$ and $K_z(t)$ for $t\in
[72,342]$ and $Re=1200$ is also reported in
fig.~\ref{fig:K_Re1000_Re1200}.  The time evolution of the total
kinetic energy shown in fig.~\ref{fig:K_Re1000_Re1200:c} presents
doubly periodic oscillations with one long period of about 8 rotation
periods and one small period of about one rotation period.  The short
period oscillations correspond to energy exchanges between the north
and south halves of the container, with a period of 2 rotation
periods. The energy exchange mechanism is visible in
fig.~\ref{fig:K_Re1000_Re1200:d} where we have reported the time
evolution of the kinetic energy of the north and south halves of the
cylinder for $t\in[312,342]$.  Similar oscillations between
north and south hemispheres have been reported to occur in a
spheroidal cavity in \cite{wu_dynamo_2009}.  

%

\begin{figure}[ht]
\centerline{
\subfigure[linear regime]{
\includegraphics[width=0.24\textwidth]{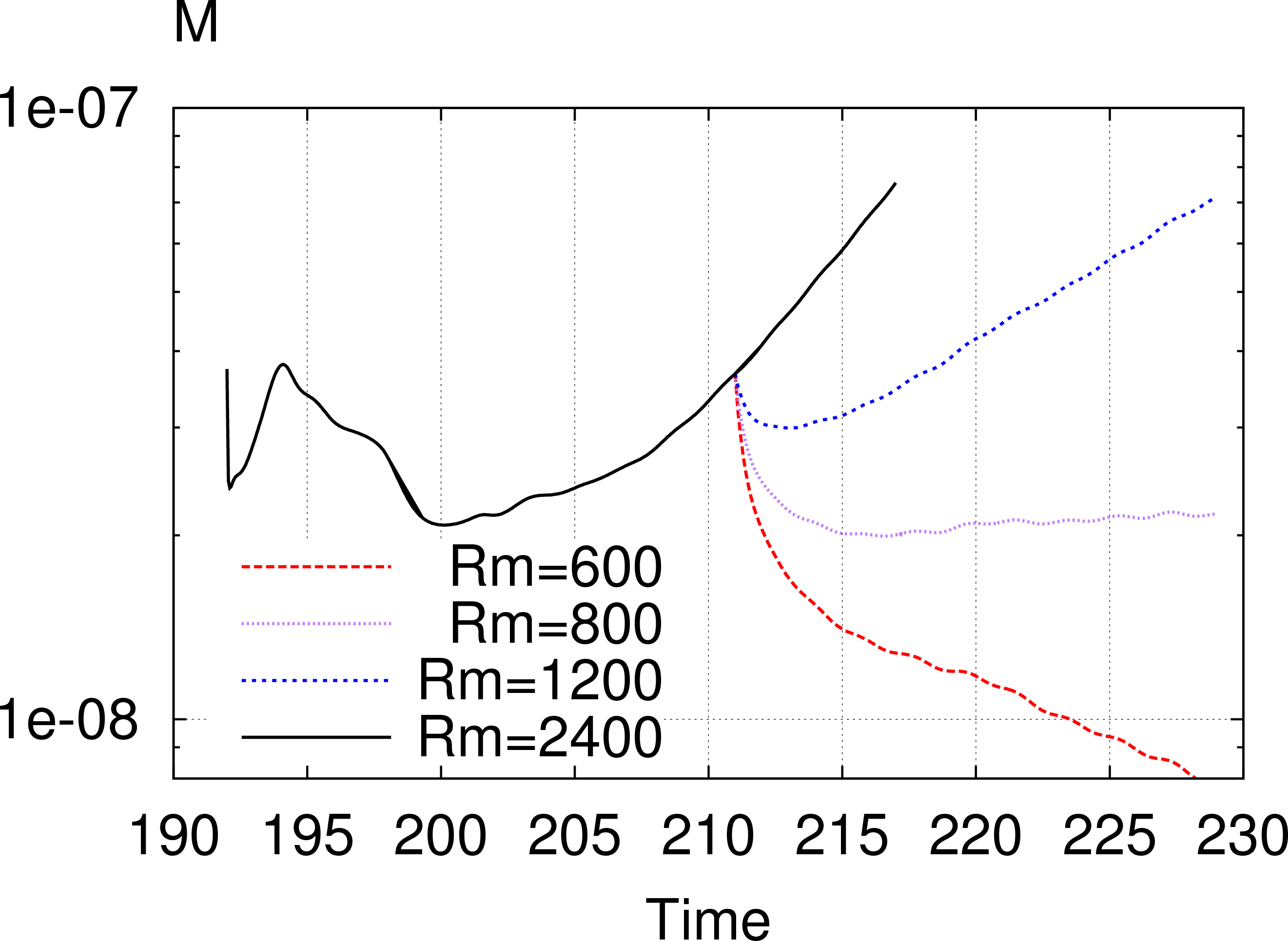}\label{fig:M_Re1200:a}}
\subfigure[nonlinear regime]{
\includegraphics[width=0.24\textwidth]{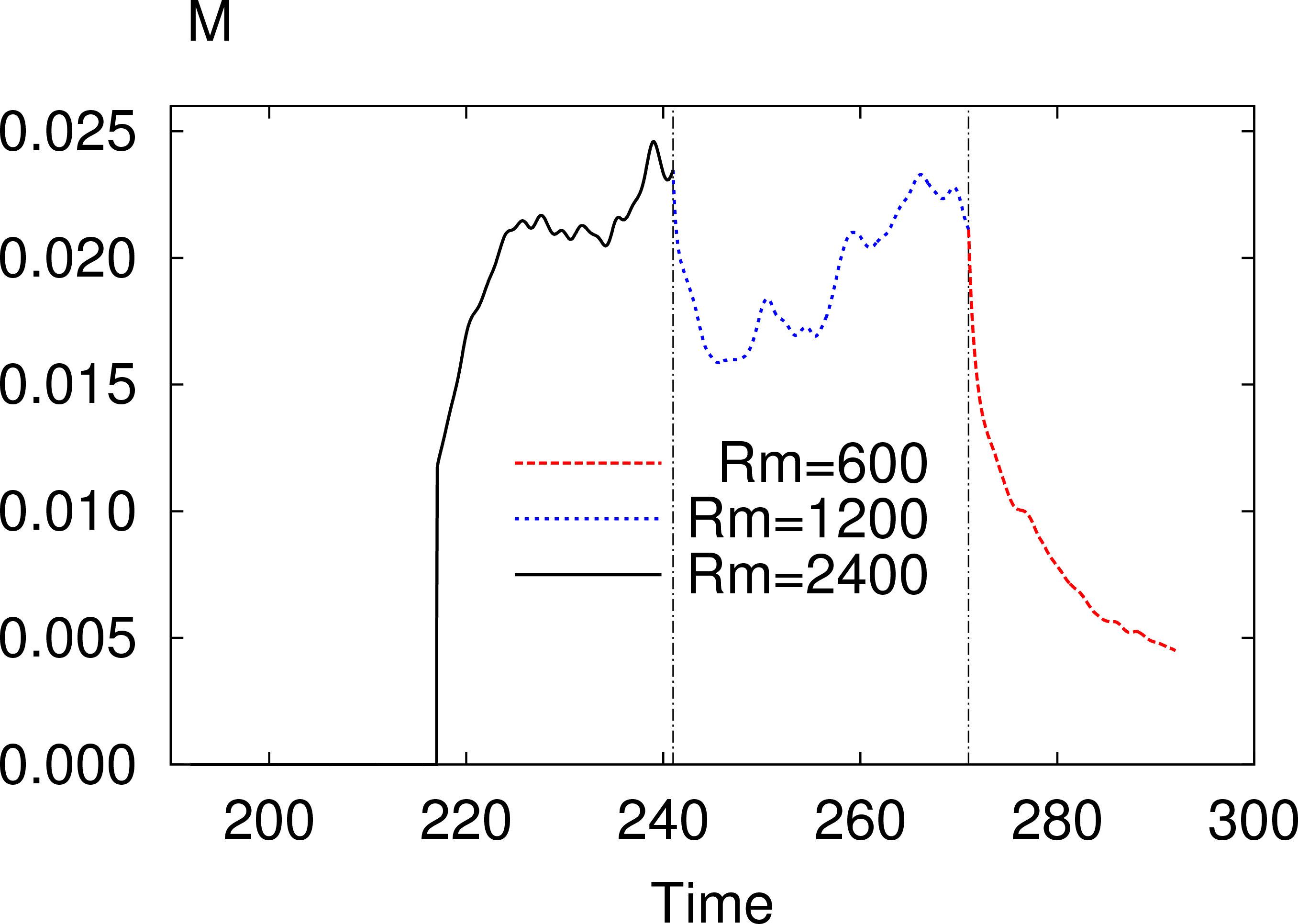}\label{fig:M_Re1200:b}}
}
\caption{Time evolution of the magnetic energy $M$ in the conducting fluid
(a) in the linear regime  from $t=192$ for $Re=1200$ and various $Rm$
as indicated (in lin-log scale) and (b)
in the nonlinear regime from $t=192$ to $t=241$ ($Re=1200, \, Rm=2400$),
from  $t=241$ to $t=271$ ($Re=1200, \, Rm=1200$)
and from $t=271$ to $t=292$ ($Re=1200, \, Rm=600$).}
\label{fig:M_Re1200}
\end{figure}

We now solve the full MHD system using as initial velocity field the
velocity computed at $t=192$ during the Navier-Stokes run at
$Re=1200$. The initial magnetic field is defined as follows to
  trigger efficiently the dynamo instability. For the Fourier modes
$m\in\{0,1\}$, the homogeneous Dirichlet boundary condition on the
scalar potential $\phi$ is replaced by $\phi=0.05 z f(t)$ for $m=0$
and $\phi=0.05 r f(t)$ for $m=1$, where
$f(t)=\frac{\tau_a^3}{1+\tau_a^3}\left(1-\frac{\tau_e^4}{1+\tau_e^4}\right)$
with $\tau_a=\frac{t}{0.4}$ and $\tau_e=\frac{t}{2}$. For $m\ge 2$,
the amplitude of each Fourier mode of the initial magnetic field
components is set to $10^{-5}$.
Various MHD runs are done at $Re=1200$ for different values of the
magnetic Prandtl number. The onset of dynamo action is monitored by
recording the time evolution of the magnetic energy in the conducting
fluid, $M(t)=\frac12 \int_{\calC} \bh^2(\br,t) d\br/K_0$.  Dynamo action
occurs when $M(t)$ is an increasing function of time for large
times. The time evolution of $M$ for $Pm\in \{2,1,\frac23,\frac12\}$
are shown in fig.~\ref{fig:M_Re1200:a}. The runs at $Pm\in
\{1,\frac23,\frac12\}$ are done by using the velocity and magnetic
fields obtained from the run $Pm=2$ at $t=211$ as initial velocity and
magnetic fields. The flow is above dynamo threshold for $Pm=1$ and
$Pm=\frac23$ but is subcritical for $Pm=\frac12$. Linear interpolation
of the growth-rates gives the critical magnetic Prandtl number
$Pm^*\approx 0.625$ corresponding to the critical magnetic Reynolds
number $Rm^*\approx 750$.

We now wish to observe the nonlinear saturation and evaluate the
impact of the magnetic Prandtl number on the nonlinear regime.  To
reach nonlinear saturation in reasonable CPU time, we have used as
initial data for the velocity and magnetic fields the velocity and
magnetic fields from the MHD run $Pm=2$ at $t=217$.  The velocity
field has been kept unchanged but we have multiplied by 300 the
amplitude of the Fourier modes $m=0,\ldots,5$ of the magnetic field.
The time evolution of the magnetic energy of this nonlinear run in the
time interval $t\in[192,241]$ is shown in
fig.~\ref{fig:M_Re1200:b}. We observe that $M$ grows smoothly until
$t=222$ and begins to oscillate thereafter. The ratio $M/K$ is
observed to be of order $10^{-2}$ during the nonlinear oscillating
regime.  After restarting the MHD run at $t=241$ with $Pm=1$ and
running it until $t=271$, we observe that the dynamo is still
active. After restarting the MHD run at $t=271$ with $Pm=\frac12$ and
running it until $t=292$, we observe that the dynamo dies in a short
time lapse, indicating that the dynamo bifurcation is not
  sub-critical for this set of control parameters.  This experiment
confirms the interval $\frac12 < Pm^* < \frac23$ for the critical
magnetic Prandtl number for dynamo action which has already been
observed in the linear regime.

Tilgner~\cite{tilgner_precession_2005} has observed that unsteadiness
and breaking of the centro-symmetry of the flow facilitate dynamo
action. A similar observation has been made in~\cite{wu_dynamo_2009},
and dynamo action is reported therein to occur when cyclic
oscillations of the kinetic energy between the north and south halves
of the spheroidal cavity occur. Although the loss of centro-symmetry is
not a necessary condition for dynamo action, we now want to test this
idea in the present cylindrical setting.

The loss of centro-symmetry of the velocity field can be monitored by
inspecting its symmetric and antisymmetric components: $\bu_s(\br,t)=
\frac{1}{2}(\bu(\br,t)-\bu(-\br,t))$ and $\bu_a(\br,t)=\frac{1}{2}
(\bu(\br,t)+\bu(-\br,t))$.  In the Navier-Stokes simulations reported
below, we monitor the loss of centro-symmetry by inspecting the time
evolution of the asymmetric kinetic energy $K_a(t)= \frac{1}{2}
\int_{\calC} \bu_a^{2}(\br,t) d \br/K_0$ and we define the asymmetry
ratio $r_a(t)=K_a(t)/K(t)$. The computations reported below have been
done on centro-symmetric grids, but centro-symmetry is not otherwise
enforced.

\begin{figure}[ht]
\centerline{
\includegraphics[width=0.4\textwidth]{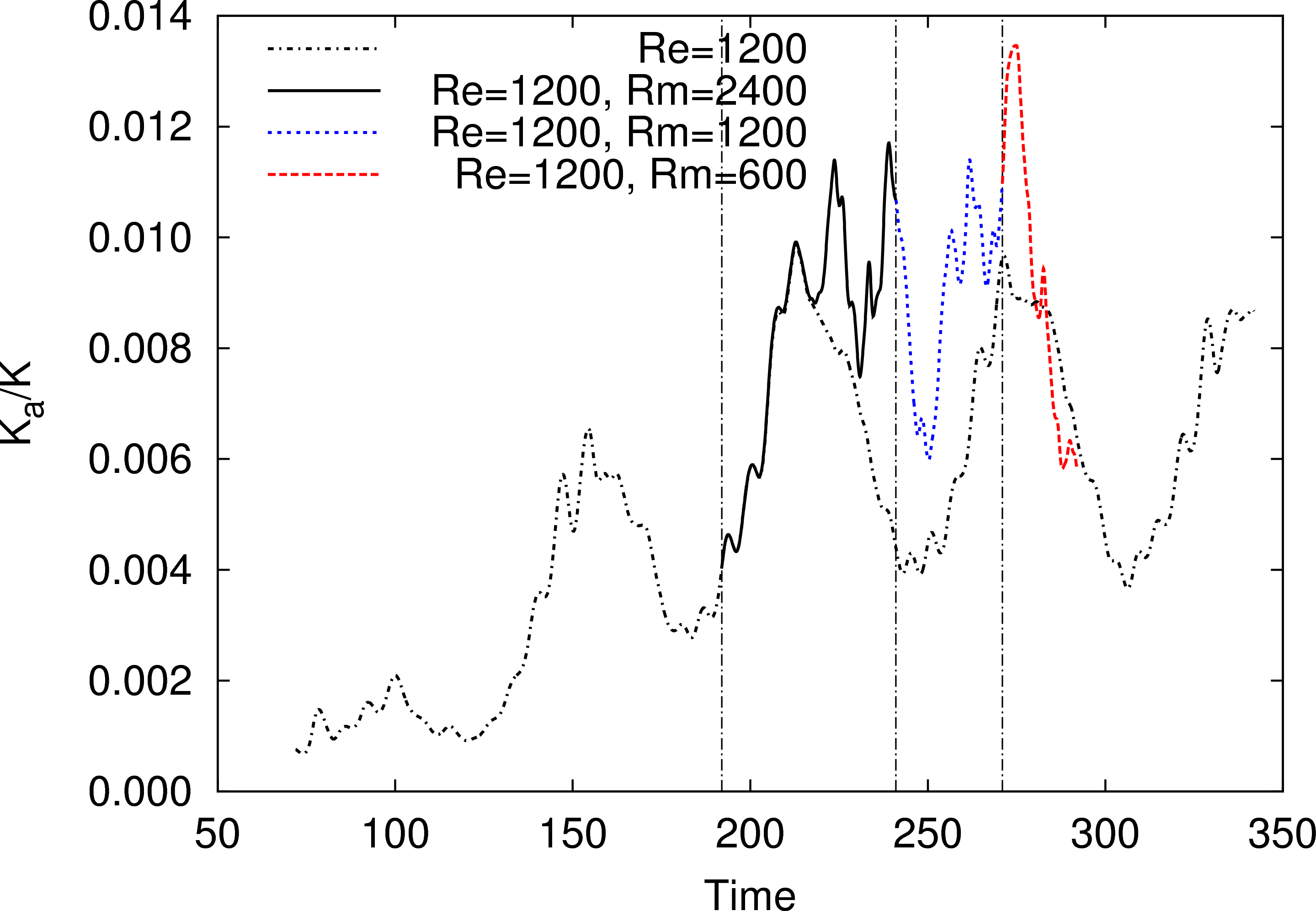}
}
\caption{Time evolution of the asymmetry ratio $r_a$ at $Re=1200$ for
  $t\in[72,342]$ in Navier-Stokes regime and $Re=1200, \, Rm=2400$ for
  $t=[192,241]$, $Re=1200, \, Rm=1200$ for $t=[241,271]$, and
  $Re=1200, \, Rm=600$ for $t=[271,292]$ in MHD regime.}
\label{fig:KasK}
\end{figure}

The time evolution of the asymmetry ratio $r_a$ is shown in
fig.~\ref{fig:KasK} for the precessing cylinder at $Re=1200$ in the
time range $t\in[72,342]$ (dotted line). The ratio $r_a$ varies
between $0.004$ and $0.01$ when the nonlinear regime is well
established, \ie $t\ge 220$. In order to evaluate the impact of the
dynamo on the symmetry of the flow, we have started the MHD run at
$t=192$ with $Pm=2$ (\ie $Rm=2400$). The time evolution of $r_a$ is
shown in solid line in fig.~\ref{fig:KasK}. Note that the solid and
dotted lines coincide since the dynamo regime is linear in the time
interval $t\in [192,217]$ and the magnetic field is too weak to have
an impact on the energy ratio $r_a$. We have restarted the MHD run at
$t=217$ after multiplying the amplitude of the magnetic field by $300$
as already mentioned.  The
ratio $r_a$ clearly departs from its Navier-Stokes value thereafter as
seen in the figure. At saturation, $r_a$ oscillates between $0.09$ and
$0.011$; these values are slightly greater than those reported
in~\cite{tilgner_precession_2005} for a precessing sphere.  We have
restarted the MHD run again at $t=241$ after reducing the value of $Pm$
to $1$, thereby reducing the magnetic Reynolds number to
$Rm=1200$. The asymmetry factor is not dramatically affected by the
change, as seen on the figure.  We have finally restarted the MHD run
at $t=271$ after reducing the value of the magnetic Prandtl number to
$\frac12$. As expected the dynamo dies and $r_a$ decreases to its
hydrodynamical level.

\begin{figure}[ht]
\centerline{
\includegraphics[width=0.33\textwidth]{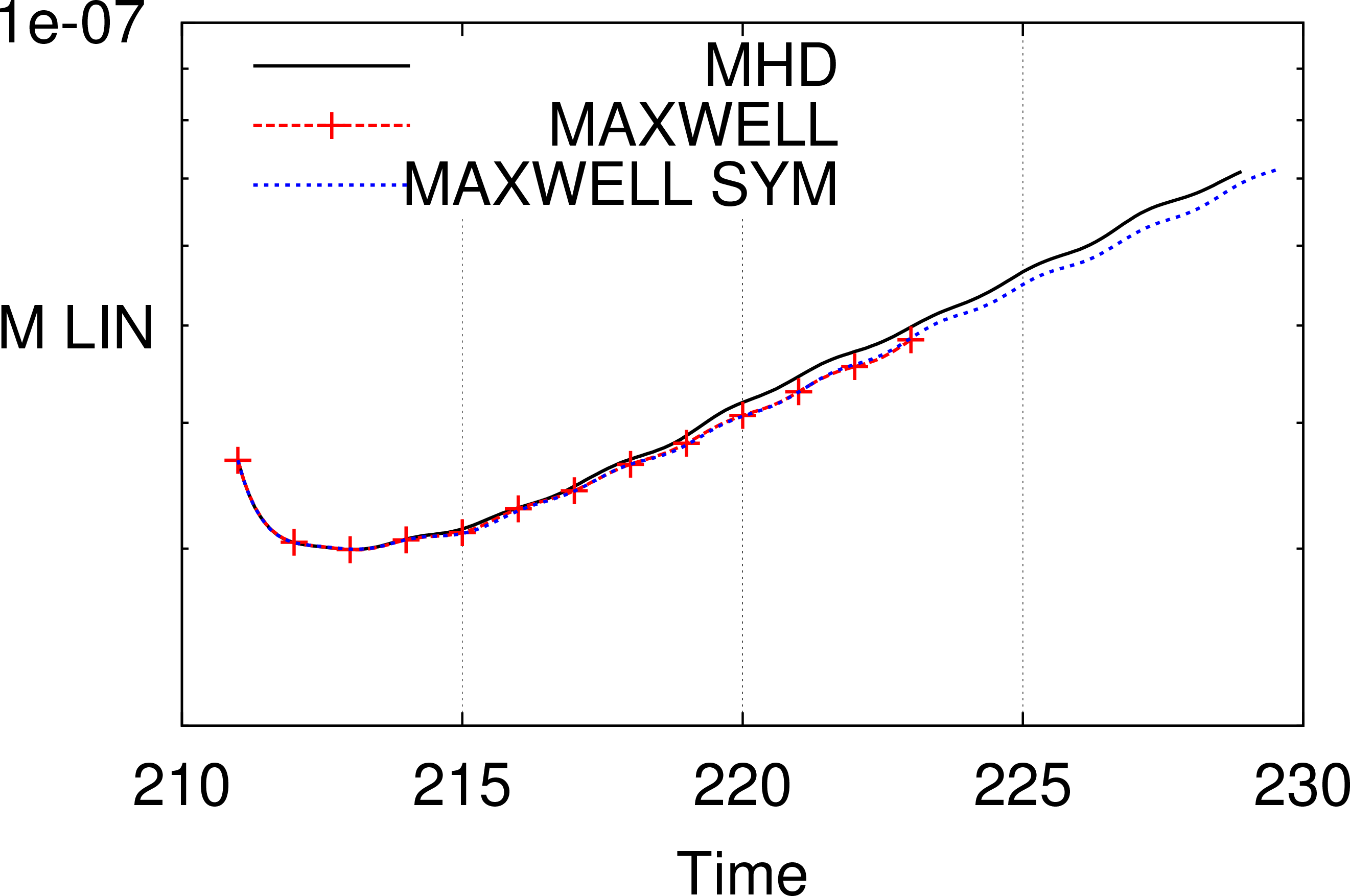}
}
\caption{Time evolution of the magnetic energy $M$ at $Re=1200$ and
  $Rm=1200$ for $t\in [211, 229]$ in MHD mode (denoted as 'MHD'), in
  Maxwell mode with the velocity frozen at $t=211$ (denoted as
  'MAXWELL') and in Maxwell mode with the symmetrized velocity frozen
  at $t=211$ (denoted as 'MAXWELL SYM').}
\label{fig:M}
\end{figure}

In order to study the impact of the centro-symmetry and the
unsteadiness of the flow on the dynamo action, we have performed two
Maxwell runs at $Rm=1200$ with the following characteristics: (i) the
velocity field at $Re=1200$ is frozen at $t=211$, (ii) the velocity
field at $Re=1200$ is frozen at $t=211$ but only its symmetric
component is retained so that the resulting velocity field is
centro-symmetric. The time evolution of the magnetic energy of the MHD
run and the two Maxwell runs (i) and (ii) are shown in
fig.~\ref{fig:M}. It is remarkable that, in the two considered kinematic
runs, the dynamo keeps growing with
a rate similar to that of the MHD run. These computations show that neither the temporal oscillations
nor the flow asymmetry play a crucial role on the dynamo action in the
precessing cylinder at $Rm=1200$.

\begin{figure}[ht]
\centerline{
\subfigure[from the side]{
\includegraphics[width=0.15\textwidth]{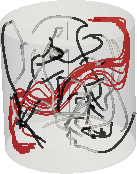}\label{fig:KasK_Re1200:a}}
\subfigure[from the top]{
\includegraphics[width=0.35\textwidth]{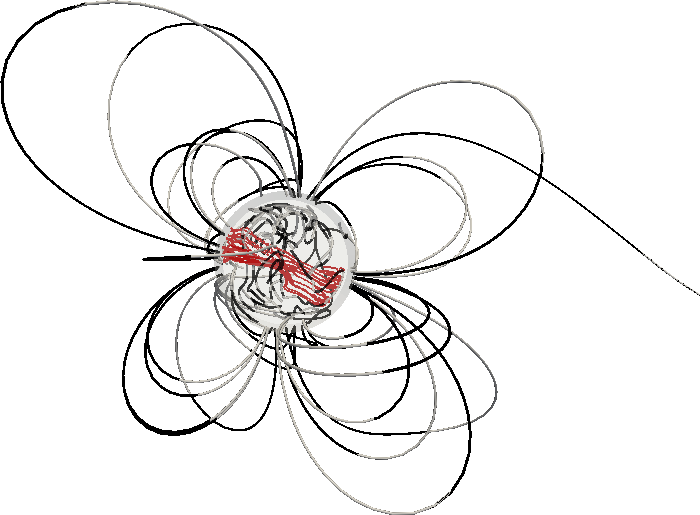}\label{fig:KasK_Re1200:b}}
}
\caption{Snapshot at $t=241$ for $Re=1200$, $Rm=2400$ showing
  vorticity field lines (red) and magnetic field lines
  colored by the axial component in the cylinder (grey/black for
  positive/negative $h_z$ component).}
\label{fig:KasK_Re1200}
\end{figure}

A snapshot of the vorticity and magnetic lines at $Re=1200$, $Rm=2400$
is shown in fig.~\ref{fig:KasK_Re1200}. We observe a central S-shaped
vortex which is deformed by the precession and reconnects at the walls
through viscous boundary layers, (see fig.~\ref{fig:KasK_Re1200:a}).  The
magnetic field lines exhibit a quadrupolar shape which is best seen in
the vacuum from the top of the cylinder (see
fig.~\ref{fig:KasK_Re1200:b}).  The magnetic energy is dominated by
azimuthal modes $m=1$, $2$, $3$.

Forty years after the promising experiments with liquid sodium by
Gans~\cite{Gans70}, we have numerically demonstrated dynamo action in
a precessing cylindrical tank.  The bifurcations through
symmetry breaking and cyclic time dependence are similar to those
already observed in dynamo flows in spherical or spheroidal precession
driven cavities.  There is however a large gap between the control
parameters used in the present simulations and those achieved in
experimental set-ups and planetary dynamos, where $E=1/Re$ and $Pm$
are many orders of magnitude smaller.  Following this preliminary
evidence for dynamo action, two further steps appear now as most
urgent: (1) studying parity breaking and unsteadiness by varying the
forcing parameters (precession angle and rate); (2) searching for a
scaling law for the critical magnetic Reynolds number as a function of
the hydrodynamic Reynolds number.  Such a relation has been proposed
by Tilgner in a precessing sphere~\cite{tilgner_precession_2005}, who
argues that it is the asymmetric part of the flow that plays a key
role in the dynamo. The research program (2) will be time consuming
as it will necessitate large scale computations to explore a wide
range of Reynolds numbers. It will also require to develop nonlinear
stabilization techniques to simulate small scale viscous
dissipation. A major step in the understanding of precession dynamo
will hopefully be achieved in the near future with the construction of
the large scale MHD facility DRESDYN at Helmholtz-Zentrum
Dresden-Rossendorf (Germany). The cooperation between simulations and experiments
will lead to a better understanding of natural dynamos,
including the geodynamo.

This work was performed using HPC resources from GENCI-IDRIS (Grant
2010-0254). We acknowledge fruitful discussions with
D.~C\'ebron, W.~Herreman, P.~Lallemand, P.~H.~Roberts, F.~Stefani and A.~Tilgner.


\bibliographystyle{plain}
\bibliography{biblio}

\end{document}